\documentclass{gistt}

\usepackage{amsmath}

\bibliography{references.bib}

\title{ADS Performance Revisited}
\author{Alexander Weber, Jobst Hildebrand, Holger Eichelberger\\
\{webera, hille003, eichelbe\}@uni-hildesheim.de\\
Institut für Informatik, Software Systems Engineering, University of Hildesheim, Hildesheim}

\begin{document}

\maketitle

\begin{abstract}

Real-time measurements are important for in-depth control of manufacturing processes, which, for modern AI methods, need integration with high-level languages. In our last SSP paper we investigated the performance of a Python and a Java-JNA based approach to integrate the Beckhoff ADS protocol for real-time edge communication into an Industry 4.0 platform. There, we have shown that while Java outperforms Python, both solutions do not meet the desired goal of 1-20kHz depending on the task. However, we are are still lacking an explanation for this result as well as an analysis of alternatives. For the first topic, we show in this paper that 1) exchanging Java-JNA with Java-JNI in this setting does not further improve the performance 2) a C++ program realizing the same behavior in a more direct integration does not perform better and 3) profiling shows that the majority of the execution is spend in ADS. For the second topic, we show that alternative uses of the ADS library allow for better performance. 

\end{abstract}

\section{Introduction}

In this paper, we revisit the integration of the TwinCAT ADS protocol into an Industry 4.0 platform as a connector. The aim of the platform is to help integrating the many different machines running in a modern shop floor and to facilitate collection and utilisation of the data generated by the different sensors on the machines. With this data, the production can be analyzed for potential faults and be supported by artificial intelligence (AI) to adapt to new situations. Machines can generate masses of real-time data during operation, which needs then to be transferred into higher level languages like Python or Java for analysis.

To allow for interoperability with many different protocols, a platform like our open source oktoflow (former IIP-Ecosphere) approach\footnote{\url{https://oktoflow.de}} provides different connectors of which the TwinCAT ADS protocol is one. The ADS protocol is a way to access data from Beckhoff PLCs that run TwinCAT\footnote{\url{https://infosys.beckhoff.com/}}. In our last SSP paper~\cite{weber2023performance} we compared an integration using the Python library pyADS and a custom Java-JNA solution based on the proprietary native ADS library by Beckhoff. We showed that the Java based approach could read data from the ADS protocol 2-3 times faster than the Python based approach, for both, local and network connection. Both approaches managed an 8 ms machine pace desirable for modern platforms~\cite{ESS22}, but fell short of an envisioned goal of up to 20kHz sampling~\cite{weber2023performance}, while the computational resources of the test device had plenty of headroom. 

Besides our work in~\cite{weber2023performance}, some literature on ADS is available: Liang et al.~built a digital twin for a robot using ADS for communication~\cite{liang2022real}. There, the average Ethernet transmission speed for 16 data points is 9.45 ms. Galeas et al.~\cite{galeas2023ethercat}~applied ADS to control a telescope via parallel connections for sending and receiving data. The authors achieved an average response time of 1 ms for over 50,000 data points. These works focus on using ADS and evaluating response times, but, in contrast to our work, the actual integrations, their limitations and alternatives are not discussed.

In this paper we further investigate the performance behavior observed and ask 1) whether there is a limiting part and 2) whether we can improve over~\cite{weber2023performance}, in particular achieve at least 10 kHz using a different approach. For this, we created a JNI-based connector and ran performance tests on it, investigated the process using a profiler and created a small C++ based program as baseline. 
We observed that neither the method of integration nor the language had an effect on the performance. But with notifications we manage to achieve a reading speed of 10kHz while encountering instability beyond this for our hardware.

\section{Approach}\label{sect-approach}

To answer our questions, we investigate whether 1) a lower level integration of the Beckhoff library through JNI performs better than JNA in~\cite{weber2023performance}, 2) profiling as well as baselining with a different programming language sheds light on root causes for performance loss and 3) we are on the wrong track, e.g., whether the ADS library itself provides faster alternatives. 

To improve the performance behavior of the ADS integration, we firstly implemented a \textbf{JNI} based solution as JNI is said to be up to one order of magnitude faster than JNA. To be comparable, we followed the general structure of the JNA solution in~\cite{weber2023performance}, used JNI instead of JNA and, instead of the proprietary native ADS library, aimed for a repeatable experiment setup using the open source \texttt{ADSToJava} library\footnote{\url{https://github.com/Beckhoff/AdsToJava}} 
by Beckhoff
, which is also available for Linux.
To be further comparable to~\cite{weber2023performance}, we also implemented our caching of the ADS memory offsets (\texttt{indexOffSets}), one reason that Java behaved faster than Python.

When the JNI-based approach did not result in the expected performance improvement we employed a \textbf{profiler} to better understand the performance behaviour of the JNI and JNA-based solutions and to find potential performance limitations in Java or ADS itself.

To further solidify the observations, we conducted an experiment with a \textbf{non-Java} program, e.g., written C++, which directly uses the ADS library. This program would have to read/write the same data as the Java variants, i.e., using the \texttt{adsSyncReadReqEx2} or \texttt{adsSyncWriteReqEx} library functions. However, in contrast, this program is only intended for experimentation rather than for a later integration into oktoflow.

Besides the ADS read and write commands that we use in heritage of~\cite{weber2023performance}, there are \textbf{ADS alternatives}, namely notifications, i.e., a specific variable can be subscribed to and a callback function gets called when the variable changes. We  expect that the notification-based approach performs best. 
In more details, notifications are enabled through an \texttt{ADSCallbackObject}, which defines a listener, a mode indicating how the listener shall be called (in our case trigger-on-change) and a maximum delay after which the callback function will be invoked anyway. A notification will deliver the affected value(s) and an \texttt{ADSNotificationAttrib} describing the size of the value(s) in memory. It is important to note that a PLC program is repeatedly executed in cycles of a given duration and that notifications per variable will be triggered only once per cycle to check if the observed value has changed. Our notification experiment uses the JNI library.

\section{Experiment}

In the experiment we investigate the performance of JNI vs. JNA as well as potential obstacles when trying to read/write values at up to 20kHz frequency. To ensure comparability we re-run the experiment from~\cite{weber2023performance} in the same conditions. We also investigate how close we could get to this goal using notifications.

The \textit{experiment setup} is using a Laptop (Intel Core i7-8665U with 4 cores and 32GB of RAM) with a constant power connection as well as a Beckhoff IPC C6930 PLC/edge device (Intel Core i7-7700, 4 CPU cores with one core isolated for real-time tasks, 32 GB RAM, Windows 10 IoT Enterprise), both connected via a Gigabit managed Ethernet switch. For simulating the Beckhoff PLC in experiments without network we use Windows 10 on the Laptop as this is required by the Beckhoff TwinCAT 3 programming environment.
For JNA, we used the proprietary native TwinCAT \texttt{TcAdsDll} library on version 2.11.0.41. 
We built \texttt{ADSToJava}\footnote{\url{https://github.com/Beckhoff/AdsToJava/tree/main}} version 3.1.0 using VS Build tools 19 and the suggested Microsoft implementation of the OpenJDK 21.0.4. We used the same Java to run the experiments. To be comparable with~\cite{weber2023performance}, we focus on the Windows version of the library rather than the Linux version of \texttt{ADSToJava}.

As \textit{experimental subjects}, we use a TwinCAT project for programming/simulating the PLC, which, as in~\cite{weber2023performance}, defines 14 variables of basic data types and 14 of corresponding array types of length 3 (as in a three-axes spindle use case). Further, we employ the JNI/JNA/C++ programs explained in Section \ref{sect-approach}. For evaluating the notifications, 

we created a special TwinCAT program with cycle time $100\mathrm{\mu}s$, which increases a \texttt{DInt} variable per cycle, thus, simulating a traceable 10kHz change rate. While we could go down to $50\mathrm{\mu}s$, the notifications begin to be unstable 
and do not deliver the data as needed.

As \textit{experimental procedure}, for the single variable read/write operations we ran 10000 read/write operations for each of the 28 variables introduced above one after another and recorded the time needed for all operations. In a local pre-experiment with the TwinCAT simulator we observed an increased variance between the datatypes when only executing 1000 or 5000 operations, which smoothed after increasing the total number of operations. 
To make sure that the comparison is fair, we re-ran the JNA based solution  with the same setting. Further, we investigated the effects of the JIT on the process, but could not identify a performance improvement with increasing runs (as we will detail in Section~\ref{sect-results}). We also observed the CPU and RAM usage throughout the pre-experiment with the Windows performance monitor and saw the CPU utilisation start a 5-6\% and raise to 6-7\% during the execution of the experiments. The RAM was stable and fluctuated by around 1GB between 5-6GB throughout each experiment. As there was no indication that the hardware is a limiting factor, we did not record these for all live experiments.

Further, we run the TwinCAT program for the notifications with 250000 changes to be able to record differences and variances. In the pre-experiments we observed randomly missing values for around 5000 changes.
We attributed the missing values to a task overload on the client side that we could fix by removing superfluous parts from the experimental subjects, e.g., a second notification on a different variable or different functions running on the same PLC cycle. This approach shows distinct CPU utilisation where it raises from a baseline of 6\% up to around 21\%. The RAM usage does change by a few hundred MB.

\section{Results}\label{sect-results}

In the experiment\footnote{\url{https://zenodo.org/records/13940333}} we investigate the performance of switching from JNA to JNI with the goal of reading/ writing values on up to 20kHz frequency. We also investigate how close we could get to this goal reading the values using notifications on change.

For the local operations, i.e., running the test programs against the TwinCAT simulation, Table \ref{table_Performance_Local} summarizes the obtained measurements for single values. The readings for arrays are not shown as they differ by only 0.019\%. We observed a similar behaviour for JNA in \cite{weber2023performance}. In total, JNA needs an average of 13.158 seconds for all types of single values where JNI requires 13.135 seconds and C++ 13.081 seconds. This shows JNI in this context is (just) 0.178\% faster than JNA and C++ is 0.58\% faster than JNA. We conclude that there is no notable difference between the three approaches and that the operations need about 30\% longer than the minimal target of 1kHz.

The remote results, i.e., when accessing the PLC via network, shown in Table~\ref{table_Performance_remote}, exhibit an average execution time for JNA, JNI and C++ of 35.110, 35.257 and 36.623 seconds, respectively. For remote experiments as well there seems to be no real difference between the approaches. They take about three times as the local experiments which is in line with the observations in~\cite{weber2023performance} for JNA.

When running both Java applications with Jprofiler\footnote{\url{https://www.ej-technologies.com/jprofiler}} and observing the methods that take the most time on the CPU, over 95\% of the time is spend on synchronous functions 

implemented by the ADS library. This explains that there is no observable difference between JNA and JNI as most of the time is spent in the library rather than in Java methods. This also fits well to the observed time behaviour of the C++ program. Further, it explains the lack of improvements through JIT as only a small portion of the runtime is executed in optimizable Java code.

While notifications do not allow for writing values, they may allow for a higher performance for read operations. We tested this on a single datatype (\texttt{DInt}) that successfully got read for all cycles. We confirmed this by recording all values and validating that all 250000 entries got read by Java.

We further observed an improvement of the local values after reducing the CPU clock in the PLC without changing the cycle times. When the read and write test are executed with a CPU clock of $100\mathrm{\mu}s$ the results get closer to 10 seconds for 10000 accesses. On local test we see averages of 10.067, 10.056 and 10.110 for JNI, JNA and C++, respectively.

\begin{table} [!htbp]
	\renewcommand{\arraystretch}{1.0}
	\caption{Results for 10000 remote accesses. 
        }\label{table_Performance_remote}
	\begin{center}
		\begin{tabular*}{1\linewidth}{@{\extracolsep{\fill}}ccccccc}
  			\hline
            \multicolumn{1}{c}{\textbf\textit{[s]}} &
			\multicolumn{1}{c}{\textbf\textit{JNA}} &
            \multicolumn{1}{c}{\textbf\textit{JNI}} &
            \multicolumn{1}{c}{\textbf\textit{C++}}
            \\
			\hline
			\hline
            \textbf{\textit{Type}} &
			\textbf{\textit{read/write}} &
			\textbf{\textit{read/write}} &
			\textbf{\textit{read/write}} & \\
			\hline
            LReal &
			36.84/31.14 & 
   			36.85/36.74 & 
            36.95/38.02
            \\
           	\hline
			LInt & 
			37.38/30.77 & 
			34.52/36.92 & 
            35.80/35.22
			\\
			\hline
			SInt & 
			30.81/37.27 & 
			37.14/33.89 & 
            36.66/36.55
			\\
			\hline
   			Byte & 
			31.04/36.98 & 
			37.63/32.19 & 
            36.36/36.00
            \\
            \hline
		\end{tabular*}
	\end{center}
\end{table} 

\begin{table} [!htbp]
	\renewcommand{\arraystretch}{1.0}
	\caption{Results for 10000 local/simulated accesses. }\label{table_Performance_Local}
	\begin{center}
		\begin{tabular*}{1\linewidth}{@{\extracolsep{\fill}}ccccccc}
  			\hline
            \multicolumn{1}{c}{\textbf\textit{[s]}} &
			\multicolumn{1}{c}{\textbf\textit{JNA}} &
            \multicolumn{1}{c}{\textbf\textit{JNI}} &
            \multicolumn{1}{c}{\textbf\textit{C++}}
            \\
			\hline
			\hline
            \textbf{\textit{Type}} &
			\textbf{\textit{read/write}} &
			\textbf{\textit{read/write}} &
			\textbf{\textit{read/write}} & \\
			\hline
            LReal &
			13.95/14.12 & 
   			15.71/14.19 & 
            15.46/14.54
            \\
           			\hline
			LInt & 
			12.60/12.56 & 
			12.73/13.32 & 
            12.02/13.16
			\\
			\hline
			SInt & 
			12.75/15.33 & 
			13.23/12.59 & 
            12.89/12.01
			\\
			\hline
   			Byte & 
			13.59/12.06 & 
			13.59/13.29 & 
            13.81/14.39
            \\
            \hline
		\end{tabular*}
	\end{center}
\end{table}

\section{Conclusion}

We investigated JNI as alternative to JNA for the integration of the ADS protocol into a Java client program as well as a C++ program and showed that the performance of the read and write operations does not differ significantly. 
We further observed the Java behaviour using a profiler and identified that over 95\% of the execution time is spent in the native library. Thus, neither the missing (expected) difference between JNA and JNI nor missing JIT optimizations are surprising. 

This also makes it unlikely that the usage of GraalVM, as suggested in~\cite{weber2023performance}, would alleviate the problem.
As an alternative for reading values we investigated the capabilities of on-change notifications showed that a sufficiently powerful machine can achieve 10kHz reading speed.

Besides reading values directly and aggregating them in higher-level programs, of course, one could pre-compute intermediary results on the PLC and achieve similar results even with synchronous reading.

In the future, we plan to integrate the open source \texttt{ADSToJava} library as model-driven connector into our oktoflow platform and aim for real application cases and further (also Linux-based) performance and energy consumption experiments in that wider context.

\printbibliography

@article{weber2023performance,
  title={Performance comparison of TwinCAT ADS for Python and Java},
  author={Weber, Alexander and Eichelberger, Holger and Schreiber, Per and Wienrich, Svenja},
  year={2023},
  publisher={Gesellschaft fuer Informatik eV},
  journal={Softwaretechnik- Trends},
  volume={43},
  number={3},
}

@inproceedings {ESS22,
author={Holger Eichelberger and Heiko Stichweh and Christian Sauer},
title={{Requirements for an AI-enabled Industry 4.0 Platform – Integrating Industrial and Scientific Views}},
_publisher={ - ThinkMind},
booktitle={Intl. Conference on Advances and Trends in Software Engineering},
year={2022},
_month={April},
pages={7-14},
isbn = {978-1-61208-946-1 / 2519-8394},
}

@article{liang2022real,
  title={Real-time state synchronization between physical construction robots and process-level digital twins},
  author={Liang, Ci-Jyun and McGee, Wes and Menassa, Carol C and Kamat, Vineet R},
  journal={Construction Robotics},
  volume={6},
  number={1},
  pages={57--73},
  year={2022},
  publisher={Springer}
}

@article{galeas2023ethercat,
  title={EtherCAT as an alternative for the next generation real-time control system for telescopes},
  author={Galeas, Patricio and Shen, Tzu-Chiang and Carrasco, Sebastian and Seguel, Rodrigo and Augsburger, Rodrigo and Huenupan, Fernando and Sep{\'u}lveda, Jorge},
  journal={Journal of Astronomical Telescopes, Instruments, and Systems},
  volume={9},
  number={1},
  pages={017001--017001},
  year={2023},
  publisher={Society of Photo-Optical Instrumentation Engineers}
}

\end{document}